# A Convolutional Attention Based Deep Network Solution for UAV Network Attack Recognition over Fading Channels and Interference

Joseanne Viana †‡, Hamed Farkhari∗†, Luis Miguel Campos ∗, Pedro Sebastião †‡,
Katerina Koutlia ¶, Sandra Lagén ¶, Luis Bernardo §‡, Rui Dinis§‡,

†ISCTE – Instituto Universitário de Lisboa, Av. das Forças Armadas, 1649-026 Lisbon, Portugal

∗PDMFC, Rua Fradesso da Silveira, n. 4, Piso 1B, 1300-609, Lisboa, Portugal

‡IT – Instituto de Telecomunicações, Av. Rovisco Pais, 1, Torre Norte, Piso 10, 1049-001 Lisboa, Portugal

§FCT – Universidade Nova de Lisboa, Monte da Caparica, 2829-516 Caparica, Portugal

¶ CTTC - Centre Tecnològic de Telecomunicacions de Catalunya (CERCA);

Emails: joseanne_cristina_viana@iscte-iul.pt, Hamed_Farkhari@iscte-iul.pt, luis.campos@pdmfc.com,
pedro.sebastiao@iscte-iul.pt, {kkoutlia, slagen}@cttc.es, rdinis@fct.unl.pt

*Abstract*—When users exchange data with Unmanned Aerial vehicles - (UAVs) over air-to-ground (A2G) wireless communication networks, they expose the link to attacks that could increase packet loss and might disrupt connectivity. For example, in emergency deliveries, losing control information (i.e data related to the UAV control communication) might result in accidents that cause UAV destruction and damage to buildings or other elements in a city. To prevent these problems, these issues must be addressed in 5G and 6G scenarios. This research offers a deep learning (DL) approach for detecting attacks in UAVs equipped with orthogonal frequency division multiplexing (OFDM) receivers on Clustered Delay Line (CDL) channels in highly complex scenarios involving authenticated terrestrial users, as well as attackers in unknown locations. We use the two observable parameters available in 5G UAV connections: the Received Signal Strength Indicator (RSSI) and the Signal to Interference plus Noise Ratio (SINR). The prospective algorithm is generalizable regarding attack identification, which does not occur during training. Further, it can identify all the attackers in the environment with 20 terrestrial users. A deeper investigation into the timing requirements for recognizing attacks show that after training, the minimum time necessary after the attack begins is 100 ms, and the minimum attack power is 2 dBm, which is the same power that the authenticated UAV uses. Our algorithm also detects moving attackers from a distance of 500 m.

## I. INTRODUCTION

Unmanned Aerial Vehicles - (UAVs) will integrate into 5G and 6G networks to provide delivery services, security, inspection, emergency services, and other functions inside and outside the network. The transportation and logistics industry will first benefit from the use of UAVs in their ecosystem, followed by all other vertical industries. In addition to coverage, high throughput, and low latency requirements, there is an increasing demand for secure and reliable connections with powerful data protection [1], [2]. We expect that emergency and high-value transportation, whose success depends on the capacity to communicate reliably and securely, will employ UAVs to provide high-quality services at lower costs [3]. Due to their aerial nature, UAVs provide faster and more flexible network services at higher data rates since they have complete control over their movement and a high probability of establishing robust line-of-sight (LOS) communication links. However, the vulnerability of wireless air-toground (A2G) communication links make UAVs susceptible to attacks that increase packet loss or, even worse, completely lose communication. In order to keep UAV communications safe, it is crucial to detect potential risks and implement countermeasures. There is extensive research on Anti-Jamming techniques. Two established approaches to identify jamming are: analyzing the packet delivery ratio and the received signal strength. However, such mechanisms deal with a high amount of lost information before detecting the attack. Moreover, in ultra-dense networks, the overall amount of connected devices might hide the presence of jammers locally. Consequently, finding other ways to address security issues in UAV networks is vital.

Currently, researchers are adopting machine learning techniques for sequence prediction problems with spatial inputs and pattern recognition [4]. Specifically, Deep Learning (DL) research exploits algorithms to make models with highlevel data abstractions by using multiple processing layers with complex structures. Especially, Deep Neural Networks (DNNs) such as Convolutional Neural Networks (CNNs) [5], [6] with Long Short-Term Memory (LSTM) or attention layers are used for temporal modeling, [7] and reducing frequency variations, respectively [6]. These characteristics make them suitable for applications that deal with time series and spatial data such as interference identification in wireless networks. The signal under analysis uses specific features to detect anomalies. Authors in [8] add attention layer in their CNN to track long temporal variations in the gradients in the time domain. There are some pre-trained networks that do not require re-design because they use transfer learning methods to learn classification procedures. The authors in [9] use pre-trained networks (i.e AlexNet, VGG-16, ResNet-50) to identify jamming using spectral images of the received signal in the UAV. However, these networks can be vast and require extensive CPU processing to sort information, making them unsuitable for use by UAVs. Despite the fact that embedded deep network techniques in the cloud or edge can monitor and evaluate channel degradation due to interference, fading and jamming attacks, anti-jamming procedures and non-traditional approaches to avoid jamming are the focus of most research on this topic, rather than recognizing attacks and there is a lack of publicly available research on attack detection in UAV communications. We intend to detect attacks against authenticated UAVs when the UAVs are providing delivery services in highly complex environments such as the realistic ones in big urban centers. We aim to add terrestrial and aerial users connected to small cells that produce interference and simulate blockages that represent the buildings. In this environment, we equip the UAV with a unique deep network design with fewer layers than pre-trained networks typically use to identify attacks.

We organize this paper as follows: Section II details the system model. It explains the channel model, the dataset, and the architecture of the intended deep network. Section III summarizes the results of the performance evaluation of the deep network. Finally, Section IV concludes this paper.

### A. Contributions and Motivation

There is a lack of research and data on the detection and prevention of jamming using deep network techniques. In order to expand the literature on this topic, we present the following key contributions of this paper:

- A comprehensive case research model that assumes interference and blockages in the scenario with authenticated UAVs in 5G networks and the presence of other UAV attackers.
- An analysis of the identification of static and moving attackers in the network with and without terrestrial users. • A smaller Convolutional Neural-Attention Based Network (CNN-Attention) architecture to detect jamming.
- Insights into the Deep network hyperparameters configuration.
- Comparison between deep network performances using attention or LSTM layers.
- Results on attack detection accuracy with and without terrestrial users in both static and moving scenarios.

Finally, we offer a visual representation of the confusion matrix for both the training and test datasets.

## II. SYSTEM MODEL

We consider a deep learning approach to detect attacks over UAV networks when there are V ($l \in \mathbb{N} \triangleq \{1,2,...,V\}$) authenticated UAVs connected to private networks in A2G links, S small cells serve U ground users, and M attackers exist with a fixed index $i \in \mathbb{N} \triangleq \{1,2,...,S\}$, $j \in \mathbb{N} \triangleq \{1,2,...,U\}$, $k \in \mathbb{N} \triangleq \{1,2,...,M\}$, respectively. The attackers are in unknown locations in the air and they can deliberately jam the signal received by the authenticated UAVs. The X-Y-Z Cartesian coordinate established between the small cells and the authenticated UAVs is defined as $||p_{bs} - p_{uav}||^2$. All the elements in the network follow small fading and fast fading propagation characteristics according to [10] and [11]. The users are in random fixed positions and they can move when the proper configuration is set up. The small cells follow the same random location positioning strategy as the users. Figure 1 illustrates a top view of of the simulation scenario. We define a total (1km x 1km) area that includes buildings with different sizes and heights, which are represented by the rectangles. The "$x$" identifies the fixed $S$ small cells available for connection. Some of the small cells are on the tops of the buildings. " • " represents the authenticated terrestrial users, "+" illustrates the attackers, and a variety of colors distinguish the authenticated UAVs from the attackers. For the sake for simplicity, the authenticated UAVs stay connected to the same base station during the entire simulation. We assume that there is sufficient space between all devices and other objects in the city in order to avoid collisions and that all devices are in outdoor locations. The small cells do not overlap coverage signals and all of the authenticated terrestrial users are always connected to the closest small cell available.

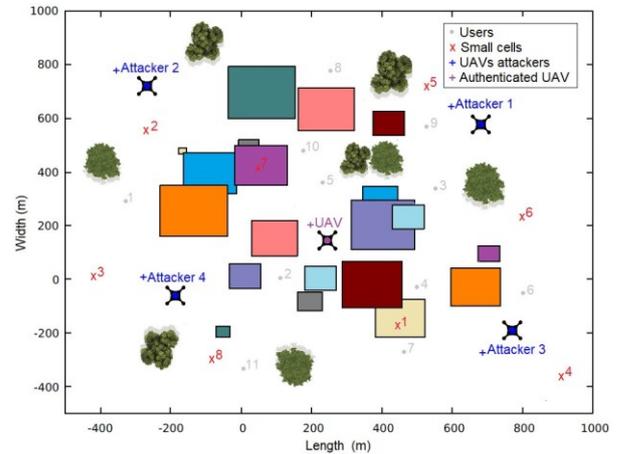

Fig. 1: Scenario Top View.

The authenticated UAVs connect to the small cells and generate downlink signals while the attackers attemp to jam the link. The jamming UAVs have the ability to adjust their power and position throughout the simulation. Their goal is to disrupt the UAV signal using decentralized capabilities and the least amount of resources possible. The attackers utilize the same propagation models as the authenticated UAVs. When they are set up with moving capabilities, they move towards the target UAVs.

TABLE I: Network Parameters.

| Scenario Parameters | Value |
| --- | --- |
| Terrestrial Users | 0,3,5,10,20 |
| Authenticated UAVs | 1 |
| Small Cells | 10 |
| Small cell height | 10 m |
| Attackers | 0,1,2,3,4 |
| Speeds | 10 m/s |
| Modulation scheme | OFDM |
| Small cell power | 4 dBm |
| Authenticated UAV power | 2 dBm |
| Attackers power | 0,2,10,20 dBm |
| Authenticated UAV position | random |
| Attackers position | random |
| Small cells position | random |
| Scenario | UMi |
| Distance | 100,200,500 m |
| Simulation time | 20 s |

We define an urban scenario for our experiment due to the complex interference and obstruction patterns that we find in a city. Additionally, this is the most common environment for the UAV emergency delivery use case. The channel between the UAVs and the small cell uses a wireless fading model which is modelled after A2G channels, and the transmission uses the OFDM modulation scheme. The deep network takes the fading and interference of wireless data into consideration using s subchannels for a total of Ni time slots, where $s \leq N_i$ (in practice, we typically have $s \ll N_i$). The urban scenarios limit the small cell's height. We use the heights and distances in Table I and we create scenarios with both LOS and non-line-of-sight (NLOS) conditions. Note that the distance in Table I is the distance between the small cell and the authenticated UAV.

### A. Channel Model

The 3GPP standards [10] and [11] describe the losses and fading in 5G UAV wireless communications. Specifically, small scale and large scale fading in rural and urban scenarios. The UAV norm adds the logarithm's height component losses to the overall calculation to differentiate the UAV's links from well-known wireless connections. Regarding the small scale fading, which is known as fast fading, there are two channel models available in the standard: the Tapped Delay Line (TDL) and the Clustered Delay Line (CDL). The second model comes from the first one. UAV fast fading models are frequently described as CDL channels. There is another subcategory for LOS and NLOS conditions in the model highlighted by the letters "ABCDE" after the name of the model. For example, the CDL-D includes line-of-sight components while the first three letters represent models with NLOS components. Due to the line-of sight characteristics of the UAV links, they are usually modelled using CDL-D. The major difference in comparison to terrestrial wireless links is that the UAV is at substantially higher altitudes considering the average rooftop height in a city, whereas, the antenna is at positions below the same reference points, which means that the angular spreads in the departure and arrival devices swap.

### B. The Dataset

As an extension of our previous work [12], we study deep network identification and generalization algorithms for jamming attacks under fading and interference when UAV attackers have static and moving configurations. We use sequential datasets, such as the time-series network parameters that the authenticated UAV generates during its mission. Specifically, we analyse two observable parameters: the Received Signal Strength Indicator (RSSI) and the Signal to Interference plus Noise Ratio (SINR) as inputs. Both parameters are collected from the authenticated UAV's receiver side. The dataset contains 2400 folders. Each folder has two files, one for RSSI data and one for SINR. The folders are classified into four configuration groups namely: *None speed*, *Attacker speed*, *User speed*, and *both speed*. The *None Speed* group collects RSSI and SINR data when there are no changes in the initial position of the elements in the network over time. In the *Attacker speed* configuration, the attackers are able to change their speed according to Table I. In the third case named *User speed*, only the users are able to move in the simulation over time and in the *Both speed* case, both elements (the attackers and the users) are able to move according to predefined speeds during the simulation time. The RSSI parameter defines the total interference power in the network and the SINR parameter measures the link quality (with the ratio of useful signal power over interference plus noise power). Both parameters are available in the authenticated UAV after the initial access synchronization.

The dataset's labels use the following nomenclature: " Yes Jamming", "No Jamming", "Moving Jamming", and "Fixed Jamming". Yes Jamming implies that at least one attacker has been discovered by the authenticated UAV in the network. No Jamming indicates the absence of jamming and suggests that the network is secure. Moving Jamming denotes that the jammer is approaching the authenticated UAV over time and Fixed Jamming suggests that the jammer is in a fixed position over time. We change the power of the jammers during the experiment.

### C. The Designed Deep Network

In the this section, we describe the deep network characteristics that recognize attacks in realistic scenarios. The motivation behind the use of a deep network solution is to learn the characteristics of networks while it is under attack. The two headed Deep Neural Network (DNN) solution receives two sequences from the observable signals RSSI and SINR and then it produces just a single classified output. The architectural design contains the following in both of the two heads: ($i$) three convolutional layers (CNN), ($ii$) a LSTM or Multi-headed-attention layer, and ($iii$) a drop-layer. The body of the deep network consists of: ($i$) three convolutional layers, ($ii$) a Drop out layer, ($iii$) a fully connected layer, and ($iv$) the output layer for two classes classification as in figure 2. After the first classification, we run another deep network with the same structure to classify the moving and non-moving jammers.

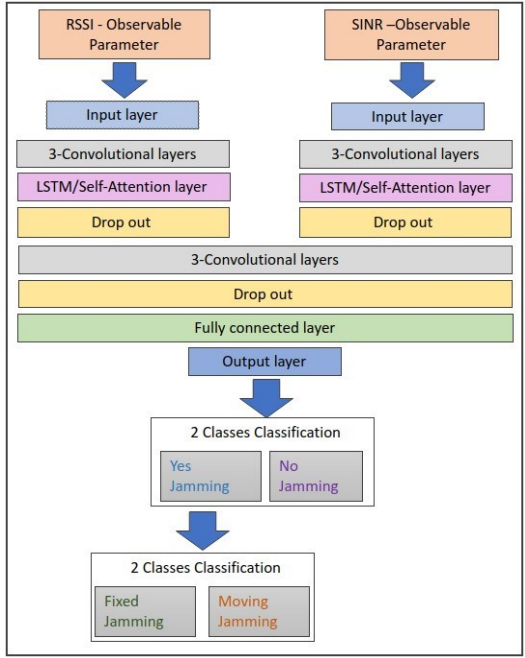

Fig. 2: Multiheaded Deep Network proposed architecture with two configurations: all structure with LSTM layer or all structure with self-attention layer

First, we use the auto correlation function to find the window size hyperparameter for the time-series data. For the other hyperparameters such as the number of CNN filters, kernel sizes, stride lengths, batch size, learning rate, and the number of regularization terms we use the grid search algorithm and other insights from our previous paper [12]. Our training data is fed into the Adam optimizer, which reduces the classification error for each new batch of commands. The Attention layer is applicable in this context because it provides the ability to capture temporal information since the nodes in the layers are weighted by the sum of the row vectors that hold the information over several time steps, which increases robustness similar to the LSTM layer, but with fewer trainable parameters. Then, we apply a technique based on Particle Swarm Optimization (PSO), Genetic Algorithms (GA), and Gradient Descent (GD) to speed up the training part of our deep network. At the end, we employ a 5-fold cross validation technique to assure correctness and prevent overfitting. The deep Network model is trained and tested on a computer system with a Nvidia RTX 3090 that has a 25Gb RAM Graphics Processing Unit (GPU). The training part for each fold took only 30 minutes using this computer system combined with the hybrid training technique. All the Convolutional layers, self-attention, and drop out groups follow the same structure and parameters mentioned in Table II. The main deep network parameters are available in Table II.

TABLE II: Deep Network Configuration Parameters.

| Deep network Parameters | Value |
|---|---|
| Base learning rate | $2.5 \times 10^{-2}$ |
| Base batch size | 32 |
| Conv-1 filters, kernel size, strides | 8, 8, 2 |
| Conv-2 filters, kernel size, strides | 8, 4, 2 |
| Conv-3 filters, kernel size, strides | 8, 3, 1 |
| Self-Attention head-number, key-dimensions | 8, 8 |
| (or LSTM) | 50 |
| Drop-out | 0.4 |
| Fully connected layer | 100 |
| Softmax | 2 |

### III. EXPERIMENTAL RESULTS

In this section, we present the results of our synthetic UAV attack dataset executed in our designed deep network. Except when explicitly mentioned, all the network and deep network parameters used are described in Table I and in Table II, respectively. For each attacker number, we ran a simulation based on the attacker power, distance, and users amount, which generated 4800 files (2400 for RSSI and 2400 for SINR). We fed this data into the deep network and we analysed the classification results. First, we calculated the overall accuracy considering all the scenarios. Our deep network was able to correctly classify approximately 84% of the scenarios regarding *Yes Jamming* and *No Jamming* labels in the training and 74% in the test.

The training results showed that the deep network miscategorized 11,407 training samples from *No Jamming* to *Yes Jamming* and vice-versa out of a total of 72288 training samples generated from all folds cross validation. For validation, we used roughly 14,000 samples. During the testing, we observed 80,537 misclassifications out of a total of 315,800. The relatively high misclassification number found in the training can be justified by taking into account the abrupt changes in the stochastic channel model and the random nature of the simulation. Moreover, the fact that we did not use samples from the same configuration in the test as the ones that we used in the training might justify the increased number of miscategorized samples.

Table 3 presents the confusion matrix for all scenarios.

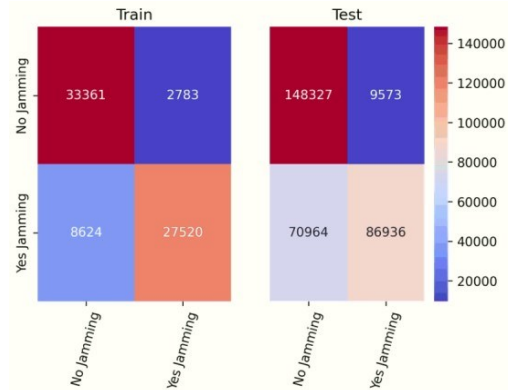

Fig. 3: The binary classification Confusion Matrix for all scenarios.

Tables III and IV illustrate additional information regarding the experiment's accuracy and f-score parameters for training and testing, respectively.

TABLE III: Precision, recall, and f1-score in training.

|  | precision | recall | f1-score | support |
|---|---|---|---|---|
| No Jamming | 0.79 | 0.92 | 0.85 | 36144 |
| Yes Jamming | 0.91 | 0.77 | 0.83 | 36144 |

TABLE IV: Precision, recall, and f1-score in testing.

|  | precision | recall | f1-score | support |
|---|---|---|---|---|
| No Jamming | 0.68 | 0.94 | 0.79 | 157900 |
| Yes Jamming | 0.90 | 0.55 | 0.68 | 157900 |

In order to simplify the 2-steps-classification in the deep network architecture in figure 2, we tried a 1-step classification with 3-classes, the labels were *No Jamming*, *Fixed jamming*, and *Moving Jamming*, but the accuracy results decreased approximately 11% reaching 72%. In the binary classification test, it became clear that the CNN layers were critical to reduce the number of trainable parameters. After replacing the LSTM with the self-attention layers, we noticed that the number of trainable parameters reduced down to approximately half of the initial amount (i.e from 43000 to 22000), but maintained the same good performance in training and validation. We observed that accuracy increased roughly by 2% (i.e 73.15 to 75.13) during testing using the attention layers. Figure 4 depicts the *accuracy* results based on the number of terrestrial users connected to the network for training and testing. The overall *accuracy* decreased according to the number of users connected to the network. For example, when there were no users in the network, training *accuracy* was about 90%, but with 20 users it was around 83%. The *accuracy* decreased because of the interference generated by the connection between the users and the small cells over time. The small fading values changed when the users were configured to move and the additional users made it hard to differentiate whether the RSSI and SINR changes were caused by attackers or users.

The accuracy of the 5-user simulation was lower (roughly 75%) compared to the other cases with fewer and/or more users because the related data was new to the deep network.

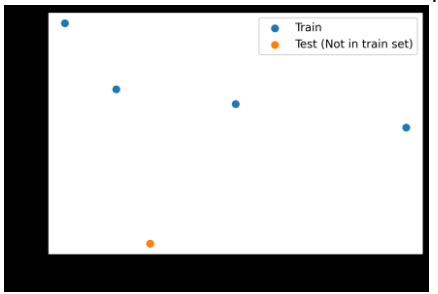

Fig. 4: Accuracy versus the number of users in the network.

These results assured the robustness of our deep network with respect to data that was not in the training. Figure 5 shows the accuracy over distance and attackers power ratios during training. Attackers with lower power are harder to be identified by the deep network. Both simulations where the attacker power was configured to 2 dBm and the distance was set up to 200 m were removed from the training.

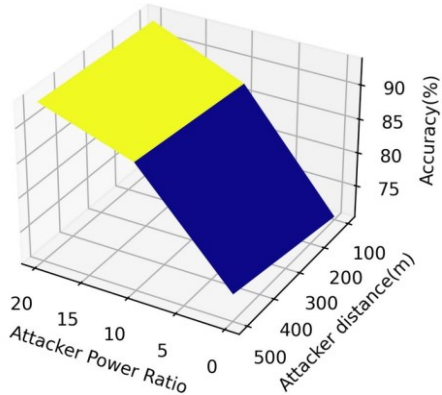

Fig. 5: Accuracy over power and distances.

Table V presents the overall results of each scenario considering all distances, powers, and attackers. The deep network achieved the highest accuracy in the Attacker speed scenario (when only the attackers move toward the authenticated UAV), but the accuracy difference compared to the other scenarios was small (i.e., 0.25% compared to None speed, 1.15% in User Speed, and 3.24 % in Both speed)

TABLE V: Accuracy when in fixed and moving scenarios.

| Scenario | Accuracy (%) |
|---|---|
| None speed | 77.35 |
| Attacker speed | 77.60 |
| User speed | 76.45 |
| Both speed | 74.36 |

Figure 6 depicts the accuracy across the number of attackers and their respective power ratios. It is difficult for the deep network to identify a small number of attackers or an attacker with limited power.

## IV. CONCLUSION

This article offered a solution based on deep networks for identifying jamming attacks in UAVs networks. We were able to embed the deep network with self-attention layer in the UAV because after training and testing the processing capacity of the generalized deep network matched the limited processing capacity of the UAV. In general, our deep network was able to recognize attacks in all scenarios' configurations. Simulations with 3 or more attackers, fewer users, and power greater than 10 were easier to be identified. Furthermore, the 3D distance between the small cell and the authenticated UAV impacted the identification accuracy. In our case, as the distance grew, the chances of identification increased because the interference decreased.

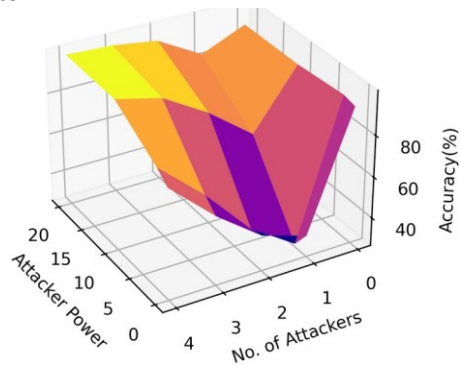

Fig. 6: Accuracy over the number of attackers in the network.

ACKNOWLEDGMENT

This research received funding from the European Union's Horizon 2020 research and innovation programme under the Marie Sklodowska-Curie Project Number 813391. documentation can be easily obtained at: